\documentclass[]{spie}  

\usepackage{amsmath,amsfonts,amssymb}
\usepackage{graphicx}
\usepackage[colorlinks=true, allcolors=blue]{hyperref}

\title{Simulation results for Robo-AO-2 using HAPA: a wavefront sensing technique for improving the adaptive optics correction of fainter stars}

\author[a]{Ruihan Zhang}
\author[a]{Christoph Baranec}
\author[b]{Marcos A. van Dam}
\author[a]{Mark R. Chun}
\author[c]{Reed Riddle}
\author[a]{James Ou}
\affil[a]{Institute for Astronomy, University of Hawaii at Manoa, Hilo, HI 96720 USA}
\affil[b]{Flat Wavefronts, Christchurch 8022, New Zealand}
\affil[c]{California Institute of Technology, Pasadena, CA 91125 USA}

\authorinfo{Further author information: (Send correspondence to R.Z.)\\R.Z.: E-mail: rzhang9@hawaii.edu, Telephone: +1 309 531 3014\\  C.B.: E-mail: baranec@hawaii.edu, Telephone: +1 808 498 9817}

\begin{document} 
\maketitle

\begin{abstract}
Direct imaging of exoplanets allows us to measure positions and chemical signatures of exoplanets. Given the limited resources for space observations where the atmosphere is absent, we want to make these measurements from the ground. However, it is difficult from the ground because it requires an adaptive optics system to provide an extremely well corrected wavefront to enable coronographic techniques. Currently only natural guide star AO systems have demonstrated the necessary wavefront correction for direct imaging of exoplanets. However, using a stellar source as the guide star for wavefront sensing limits the number of exoplanet systems we can directly image because it requires a relatively bright V$\sim$10 mag star. To increase the number of observable targets, we need to push the limit of natural guide stars to fainter magnitudes with high Strehl ratio correction. We propose to combine laser guide star (LGS) and natural guide star (NGS) wavefront sensing to achieve the high Strehl correction with fainter natural guide stars. We call this approach Hybrid Atmospheric Phase Analysis (HAPA); ‘hapa’ in Hawaiian means ‘half’ or ‘of mixed ethnic heritage’. The relatively bright LGS is used for higher order correction, whereas the NGS is used for high accuracy lower order correction. We focus on demonstrating this approach using Robo-AO-2 at the UH 2.2m telescope on Maunakea with a UV Rayleigh laser at 355 nm. The laser focuses at 10 km altitude and has an equivalent magnitude of $m_{U}$ $\approx$ 8. In this report specifically, we present simulated results of HAPA employed at Robo-AO-2, with the LGS system having a single configuration of 16x16 subaperture Shack-Hartmann wavefront sensor and the NGS system having 6 different configurations -- 16x16, 8x8, 5x5, 4x4, 2x2 and 1x1. We also discuss the on-sky experiments we plan to carry out with  HAPA at the UH 2.2m telescope.
\end{abstract}


\section{INTRODUCTION}
\label{sec:intro} 

The study of exoplanets has bloomed over the past few decades since the first discovery of a planet orbiting a sun-like star outside of our solar system, 51 Peg b \cite{Mayor1995}. Since then, most exoplanets have been discovered using radial velocity (RV) or transit curves methods. Only $\sim$ 80 exoplanets were discovered using direct imaging, which is $\sim$1$\%$ of all exoplanets discovered according to NASA. The reason for this is that imaging dim and cool planets orbiting stars that are respectively bright and hot with small angular separations is extremely difficult. High resolution and high contrast imaging is needed to perform such tasks. Nonetheless, direct imaging is an attractive method for observing exoplanets because direct imaging can provide high accuracy astrometry of exoplanets, which has encoded information of companion separation, orbital eccentricity, and companion to star mass ratio; and it opens doors to exoplanet emission and reflection spectroscopy. Spectra from exoplanets allow us to better study their atmospheric composition and further explore their habitability.

Currently, direct imaging of exoplanets is enabled by two major techniques: coronagraphs and adaptive optics (AO). Coronagraphs block out the light from the relatively bright stars in the centers of planetary systems, allowing the dim exoplanet companions to show. Ground-based direct imaging requires adaptive optics (AO) systems to achieve the necessary spatial resolution. This is because the ever turbulent atmosphere on earth works as a variable lens that distorts the incoming light from celestial objects. AO systems measure the atmospheric turbulence using wavefront sensors (WFSs) and compensate for the effect of the atmosphere using deformable mirrors; this results in near-perfect, diffraction limited images. 

Despite its numerous advantages and improvements for ground-based observations, AO has its limitations. One strong limitation lies in the method of wavefront sensing. The two most common ways to measure how the atmosphere distorts the incoming light are (1) to use a bright natural guide star (NGS) or (2) to use a laser guide star (LGS)\cite{Davies2012}, which is an artificial star created with a laser beam. The point source objects that can be used as guide stars in an AO system need to be bright enough to provide an accurate measurement of the wavefront at kHz frequencies, but there are only a sparse number of bright stars in the night sky. Therefore, NGS systems are limited by the targets they can observe. On the other hand, LGS systems have a larger sky coverage, because brightness requirements on the stellar sources for low-order wavefront sensing are much more relaxed. However, LGS systems are limited by the cone effect, also known as focal anisoplanatism. The cone effect is caused by the fact that LGSs are only ~10-90 km away from the telescope, which samples a volume in the atmosphere that is cone shaped compared to the cylindrical shape that is covered by the target from far away distances. This creates inaccuracy in the measured wavefront called focal anisoplanatism. First, the higher altitude atmospheric turbulence cannot be fully captured by a LGS. Second, the turbulence at higher altitudes gets projected onto the telescope pupil, magnifying the spatial scale of the turbulence due to the diverging rays coming from a laser guide star. 

For direct imaging of exoplanets, extreme AO is required to produce the necessary image quality. Current exoplanet direct imaging systems such as SCExAO at Subaru \cite{Martinache2009}, PALM-3000 at Palomar Observatory \cite{Dekany2013}, GPI at Gemini South \cite{Macintosh2014}, and SPHERE at the Very Large Telescope (VLT) \cite{Beuzit2019} all use natural guide star wavefront sensing. Because the number of targets observable are very limited with using natural guide stars, we want to push the natural guide star brightness limit to fainter magnitudes to increase the number of objects observable with direct imaging. We propose a wavefront sensing method called HAPA: Hybrid Atmospheric Phase Analysis, which uses LGS for high-order correction and NGS for lower order correction. ‘Hapa’ in Hawaiian means ‘half’ or ‘of mixed ethnic heritage’\cite{hapaHawaiian}, alluding to the fact that our proposed wavefront sensing method uses both NGS and LGS. Because the natural guide star is only used for lower order correction, the amount of light needed for detecting the lower order aberrations is largely reduced. This relaxes the brightness limit on the stars that can be used for NGS, and thus increases the number of targets we can observe using this technique. The idea was first introduced by Ellerbroek in 1994\cite{Ellerbroek1994}, then he successfully demonstrated it with Rhoadarmer at Starfire Optical Range in 1997\cite{Ellerbroek1997}, but there was no further pursuit after the brief demonstration.

Another commonly considered approach to push for high Strehl ratio correction to fainter stellar guide limits is laser tomography adaptive optics (LTAO), which uses multiple laser guide stars to achieve high Strehl ratio correction. Having multiple lasers allows the systems to identify what turbulence resides in which atmospheric layer and completely sample the cylindrical volume in the atmosphere, which helps largely reduce the focal anisoplanatism. Several large observatories like Keck \cite{Wizinowich2022} and Subaru \cite{Akiyama2020} are in the process of deploying LTAO systems. The major advantage to LTAO compared to NGS systems is its large sky coverage, because it is only limited by the tip/tilt guide star that goes to much fainter magnitudes. On the other hand, the disadvantage to LTAO is that current demonstrations have not been able to achieve their simulated performance and people have not been able to explain why. 
In addition, LTAO systems are expensive and complicated to build and implement, which might not be the most suitable system for budget limited observatories. Therefore, we want to explore the performance of HAPA, see how its performance compares to the laser tomography approach, and understand how it may benefit the other observatories. Given that many observatories already have AO systems that use either LGS wavefront sensing, or NGS wavefront sensing, or both, deploying a HAPA system might be easy and economical with a significant improvement in image quality.

\begin{figure}
    \centering
    \includegraphics[width=0.8\linewidth]{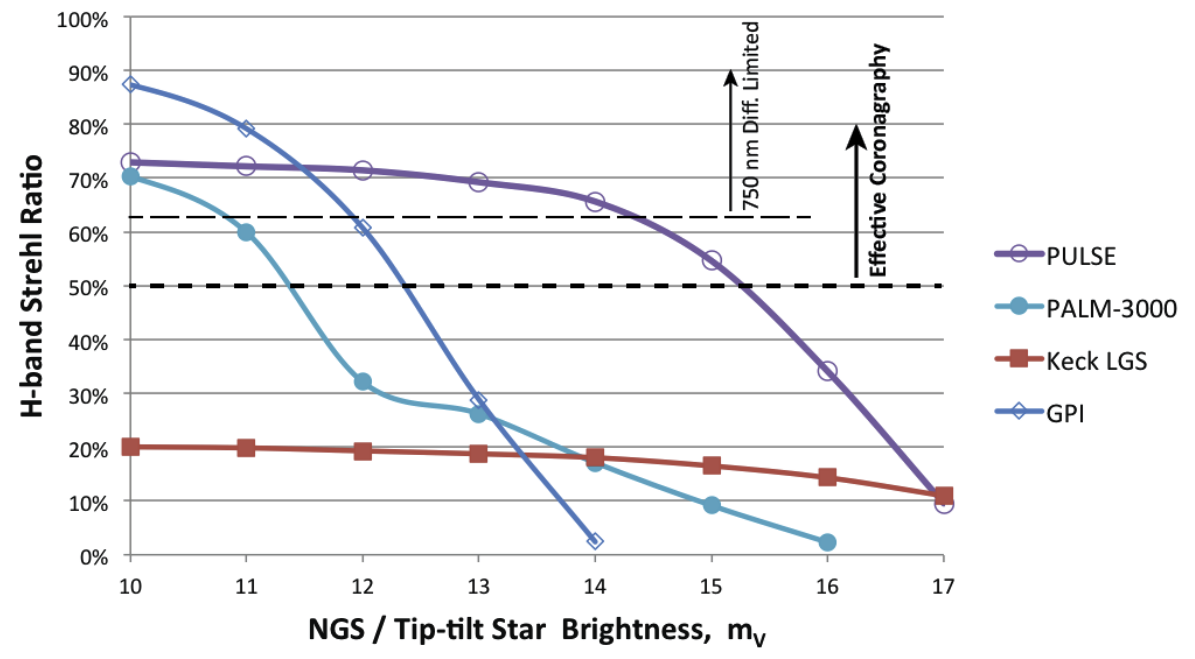}
    \caption{H-band Strehl ratio curves for multiple AO systems. The simulations show that PULSE strongly enhances the performance of PALM-3000 out to much fainter magnitudes, out performs GPI in magnitude space, and out performs the Keck LGS system in Strehl ratio. All curves are for median seeing at their respective sites. Image credit: Baranec et al. 2014\cite{Baranec2014a}.}
    \label{fig:PULSE}
\end{figure}

\begin{figure}
    \centering
    \includegraphics[width=0.7\linewidth]{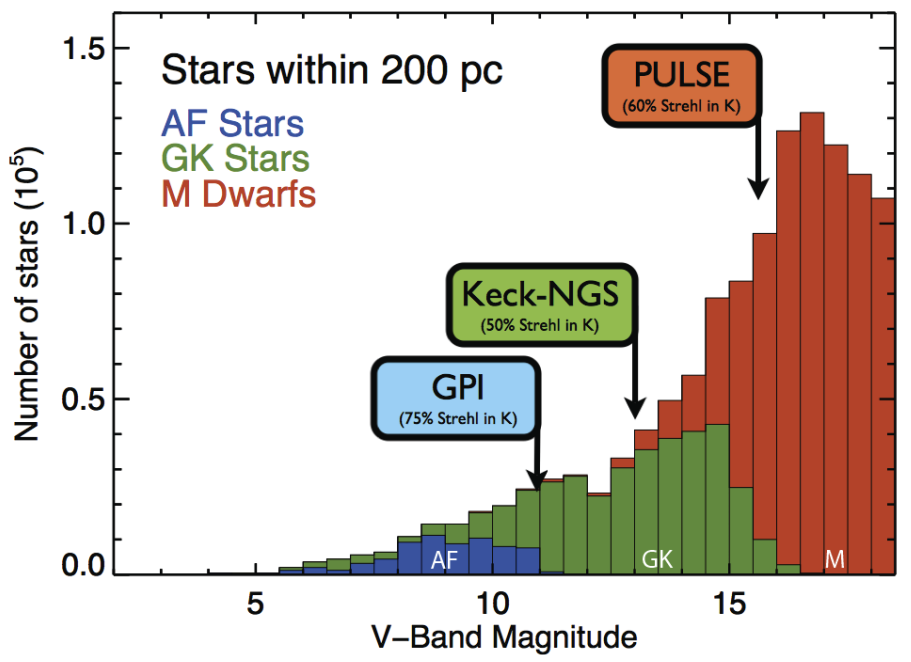}
    \caption{Stellar distribution as a function of V-band magnitude based on the TRILEGAL galactic model\cite{TRILEGAL} within 200 pc. PULSE opens new parameter space for direct imaging of exoplanets compared to the Keck NGS system and GPI. Image credit: Baranec et al. 2014\cite{Baranec2014a}.}
    \label{fig:science}
\end{figure}

HAPA initially started as the Palomar Ultraviolet Laser for the Study of Exoplanets (PULSE) intending to expand the science reach of PALM-3000\cite{Baranec2014a}. Simulations have shown that for PALM-3000, a high-contrast extreme AO system, HAPA can dramatically increase the number of observable targets for near-infrared high-contrast and diffraction-limited visible light imaging by expanding the limiting natural guide star magnitude from m$_V < 10$ to m$_V < 16$ (Figure \ref{fig:PULSE}). The PULSE simulations assumed a LGS system that has a 40 x 40 subaperture wavefront sensor running at 1 kHz frame rate and a NGS system with a 8 x 8 low-order wavefront sensor capable of running at three frame rates of 1 kHz, 250 Hz, and 100 Hz. The PULSE performance curve shown in Figure \ref{fig:PULSE} is optimized with the three selectable frame rates. As the natural guide star magnitude gets fainter, the wavefront sensor frame rate slows down to adapt for the number of photons collected and gets measurements with reasonable signal to noise ratios. PULSE opens up direct imaging opportunities to study exoplanets around M dwarfs and young stars of earlier type stars, which are too dim for current direct imaging AO systems (Figure \ref{fig:science}). Unfortunately, HAPA/PULSE did not come to realization with PALM-3000 due to limited resources, but it is still an attractive wavefront sensing method that may benefit the AO community. Therefore, we are continuing the exploration and investigation of HAPA using Robo-AO-2\cite{Baranec2018}, a robotic AO instrument deployed at the UH 2.2m telescope (UH2.2) on the summit of Maunakea. Even though HAPA does not serve strong advantages for sciences conducted at Robo-AO-2, demonstrating the method on-sky as a proof of concept and using the results to understand how realistic our simulations are is an important milestone to develop a novel wavefront sensing technique.

We aim to gain a detailed understanding of HAPA: how much it improves AO correction for traditional single conjugate AO systems and how it may benefit differently sized telescopes with different AO systems. We plan to carry out on-sky experiments for HAPA using Robo-AO-2, which was designed and built with performing experiments for HAPA in mind as one of the science goals, it has many special built-in mechanisms. For example, both the laser and stellar wavefront sensing system use the same computer, which makes the integration process much easier. Both wavefront sensor cameras have variable frame rates. The stellar wavefront sensing system has multiple lenslet arrays that can be exchanged quickly to allow for changing the spatial frequency of sensing. There are also neutral density filters in a filter wheel that can be used to quickly change the brightness of the stellar guide star without having to slew to different targets (with the understanding that sky background is similarly attenuated). All these built-in mechanisms simplify the experiment procedure for demonstrating HAPA and largely reduce the telescope time we need on-sky. Before carrying out the on-sky experiments, we need to run AO simulations for HAPA to understand how HAPA may improve Robo-AO-2 performance at UH2.2 under various weather circumstances and thus help us design the on-sky experiments. This work reports the current progress of the HAPA project at its simulation stage. Section \ref{sec:repOldResults} presents past simulations that were explored during the instrument design phase of Robo-AO-2; Section \ref{sec:currentResults} showcases simulation results of the as-built system; and Section \ref{sec:expPlans} discusses plans for the on-sky experiments with Robo-AO-2 at UH2.2.

\section{Replicating Previous Results}
\label{sec:repOldResults}
During the design phase of Robo-AO-2, van Dam and Baranec ran simulations to demonstrate HAPA's performance with Robo-AO-2 on UH2.2 at the summit of Maunakea\cite{Baranec2018,vanDam2016}. The simulations are written in $yao$, an adaptive optics simulation package based in yorick \cite{rigaut1a2013simulating}. As an exercise to get familiarized with the simulations and better understand the simulation parameters, we replicated old HAPA results from van Dam and Baranec (2016) \cite{vanDam2016}. This also allowed us to explore simulation errors a little bit and modify how we ran the other simulations. The simulation parameters are listed in Section \ref{subsec:oldParams} while the results are presented in Section \ref{subsec:oldResults}.

\subsection{Old simulation parameters}
\label{subsec:oldParams}
This subsection is copied from van Dam's 2016 report with some modifications: \cite{vanDam2016}

The photometry of both the guide star and the sky depends on the passband chosen. For the NGS, we use the photometric parameters for the GMT scaled to the UH2.2 telescope. The passband is assumed to be from 0.64 to 0.90 microns, which is approximated as comprising half of R-band plus the whole of I-band. We also assume that for a typical guide star, m$_{RI}$ is 0.68, so the I-band magnitude is lower than the corresponding I-band magnitude. The photometric zero point is thus 6.9×10$^{10}$. The photon return from the LGS is assumed to be 1.45×10$^7$ photodetections per second across the whole frame. The relevant
photometric and noise parameters are tabulated in Table \ref{tab:noiseParams}.

\begin{table}[ht]
    \centering
    \begin{tabular}{|l|r|r|}
        \hline
        Parameter & LGS & NGS \\
        \hline\hline
        Central wavelength & 355 nm & 790 nm \\ 
        \hline
        Bandwidth & & 150 nm\\
        \hline
        Photometric zero point & & 6.9x10$^{10}$\\
        \hline
        Sky background & & 18.5\\ 
        \hline 
        Quantum efficiency & & 0.80\\
        \hline
        Optical throughput & & 0.41\\
        \hline
        Excess noise factor & 1.45 & 1.45\\
        \hline
        Read-out noise & 0.5 e$^-$ & 0.5 e$^-$\\
        \hline
        Dark current & 1 e$^-$/s & 0.05 e$^-$/s\\
        \hline
    \end{tabular}
    \caption{Photometry and noise parameters used in the simulations.}
    \label{tab:noiseParams}
\end{table}

The LGS WFS has 19x19 subapertures with 6x6 0.75” pixels per subaperture. The LGS itself is modeled as a Gaussian spot with a FWHM of 1.8” and set at 10 km altitude. The NGS WFS can be configured with a variable number of subapertures, ranging from 1x1 to 8x8 subapertures across the pupil. Each subaperture has 4x4 0.6” pixels.
There is 20x20 actuator common-path DM, as well as a tip-tilt mirror which is common to the NGS WFS and the science field, but is not seen by the LGS WFS. The simulation uses a minimum-variance tomographic reconstructor that combines information from both the LGS and the NGS \cite{vanDam2021}. 

The maximum frame rate for the WFSs is 1500 Hz. 3000 iterations were run for a total of 2 seconds of integration time unless specified otherwise. The frame rate of the NGS WFS was optimized for each magnitude by varying it from 23 Hz to 1500 Hz. The strehl ratios are evaluated at 620 nm, which correspond to r-band in the visible. Simulations were run for a typical turbulence profile. The turbulence parameters are tabulated in Table \ref{tab:50turb}. The value of r$_0$ at 500 nm is 0.128 m. No dome seeing, vibrations, nor wind shake were included in the simulations, and these sources of tip-tilt error should be characterized.

\begin{table}[h]
    \centering
    \begin{tabular}{|l||r|r|r|r|r|r|r|r|r|r|}
        \hline
        Elevation (m) & 0 & 5 & 15 & 37 & 197 & 357 & 804 & 3171 & 7307 & 14636 \\
        \hline
        Turbulence fraction & 0.25 & 0.365 & 0.098 & 0.059 & 0.022 & 0.006 & 0.045 & 0.041 & 0.066 & 0.048\\
        \hline
        Wind speed (m/s) & 6.8 & 6.8 & 6.8 & 6.8 & 6.8 & 6.8 & 6.9 & 8.2 & 21.2 & 27.6\\
        \hline
        Wind direction (deg) & 0 & 76 & 92 & 190 & 255 & 270 & 350 & 17 & 29 & 66\\
        \hline
    \end{tabular}
    \caption{Typical turbulence profile used in the simulations.}
    \label{tab:50turb}
\end{table}

\subsection{Replicated results}
\label{subsec:oldResults}
The old results in van Dam and Baranec (2016)\cite{vanDam2016} are tabulated in Table \ref{tab:oldResults}. The LGS-only and the NGS-only results from the same study are also shown for comparison and completeness. The LGS-only results were obtained by setting the NGS system to 1x1 subaperture which performs as a tip-tilt guide star. Simulations for NGS-only AO systems were performed for 16x16 and 8x8 subapertures. 

\begin{table}[]
    \scriptsize
    \setlength{\tabcolsep}{5.5pt}
    \centering
    \begin{tabular}{|c||r|r|r|r|r|r|r|r|r|r|r|r|r|r|r|}
        \hline
         Mag&0&6&7&8&9&10&11&12&13&14&15&16&17&18&19 \\
         \hline
         \hline
         HAPA&0.695&&&0.650&0.596&0.532&0.446&0.367&0.239&0.192&0.122&0.082&0.060&0.038&0.022\\
         \hline
         NGS & 0.661&0.618&0.587&0.540&0.491&0.389&0.233&0.154&0.080&&&&&&\\
         \hline
         LGS & 0.175&&&0.174&0.175&0.174&0.170&0.164&0.158&0.134&0.118&0.082&0.060&0.038&0.022\\
         \hline
    \end{tabular}
    \caption{Best performing HAPA, LGS-only, and NGS-only strehls/subaperture number from van Dam and Baranec (2016) \cite{vanDam2016} as a function of magnitude.}
    \label{tab:oldResults}
\end{table}

\begin{figure}
    \centering
    \includegraphics[width=0.8\linewidth]{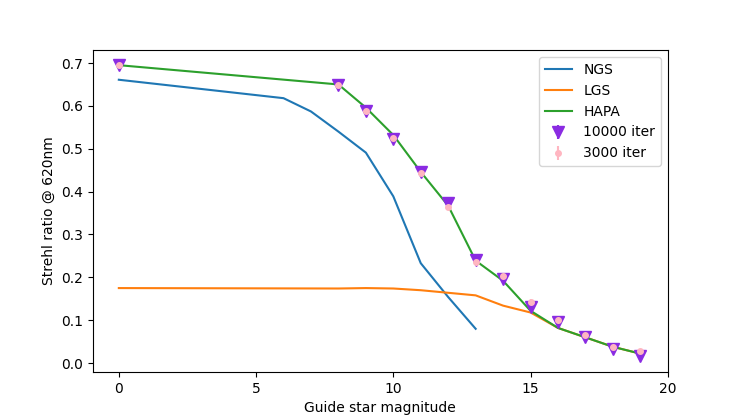}
    \caption{Simulated performances of NGS, LGS, and HAPA at UH2.2 with an initial design of Robo-AO-2. All solid lines are old results presented in van Dam and Baranec (2016)\cite{vanDam2016} and Baranec et al. (2018)\cite{Baranec2018}. The dots and the triangles with error bars are repeated results for HAPA.}
    \label{fig:repOldResults}
\end{figure}

Using exactly the same parameters as before, we were able to replicate HAPA results generated during the design phase of Robo-AO-2 (Table \ref{tab:replicatedResults}). All results including HAPA, NGS, and LGS strehls from van Dam and Baranec (2016)\cite{vanDam2016}, and the replicated HAPA results are plotted against each other in Figure \ref{fig:repOldResults} for comparison. We investigated the effect that the number of iteration cycles has on the resulting Strehl ratio. To do so, we ran simulations with 2 different number of iterations: 3000 iterations (2 s of integration time) and 10000 iterations (6.67 s of integration) while keeping every other parameter the same. 10000 iterations is the number of simulation cycles recommended in the manual of $yao$. Simulations for each data point were ran repeatedly for 30 times with each iteration number. There are slight discrepancies between each run because the simulated measurement noise is a random process. Their means were taken and we also calculated their corresponding standard deviations (STD).  

\begin{table}[h]
    \centering
    \setlength{\tabcolsep}{3.5pt}
    \begin{tabular}{|c||r|r|r|r|r|r|r|r|r|r|r|r|r|}
         \hline
         Mag&0&8&9&10&11&12&13&14&15&16&17&18&19 \\
         \hline
         \hline
         10000-iteration&0.696&0.650&0.587&0.523&0.447&0.374&0.241&0.196&0.131&0.096&0.062&0.034&0.016\\
         $\pm$STD&0.000&0.001&0.002&0.002&0.002&0.004&0.003&0.003&0.004&0.004&0.003&0.002&0.003\\
         \hline
         3000-iteration&0.695&0.649&0.588&0.524&0.444&0.364&0.237&0.204&0.143&0.100&0.065&0.038&0.027\\
         $\pm$STD&0.000&0.002&0.005&0.004&0.006&0.004&0.006&0.006&0.006&0.006&0.006&0.006&0.005\\
        \hline
        NGS configuration& 8x8&8x8&8x8&8x8&5x5&5x5&5x5&2x2&2x2&1x1&1x1&1x1&1x1\\
         \hline
         NGS frame rate [Hz]& 1500&1500&1500&750&750&375&375&188&94&94&47&47&23\\
         \hline
    \end{tabular}
    \caption{Replicated optimal HAPA results at each natural guide star magnitude with the corresponding NGS configuration and frame rate.}
    \label{tab:replicatedResults}
\end{table}

Based on the results, the number of iterations per simulation does not have a large effect on the Strehl ratio. Therefore, simulations performed from then on all used 3000 iteration cycles. The simulation results also had very small variations, which are reflected in the small STD values that are not even visible as error bars in Figure \ref{fig:repOldResults}. For bright guide stars where the NGS is the limiting factor of performance, HAPA seems to outperform the NGS-only AO system because the LGS system has 19x19 subapertures across the pupil whereas the NGS-only mode has 16x16 subapertures. This leads to a larger fitting error for the NGS-only scenario given the slightly more coarse sampling.

\section{HAPA results for the as-built Robo-AO-2}
\label{sec:currentResults}
Robo-AO-2 was installed at the UH2.2 in May of 2023 \cite{Baranec2024} and is still in the process of getting fully robotized. Similar to all other instrumentation projects, the final product always differ from the initial concept of the instrument formulated during the proposal/design phase. Therefore, we reran HAPA simulations for the as-built Robo-AO-2. We present the simulation parameters we use in Section \ref{subsec:simParamRobo-AO-2} and our results in Section \ref{subsec:simResultsRobo-AO-2}.

\subsection{Simulation parameters}
\label{subsec:simParamRobo-AO-2}
Robo-AO-2 uses a 15 W UV Raleigh laser at 355 nm to create the laser guide star at the altitude of 10 km. The UV LGS has an intrinsic kernel size of 1.2", which gets imaged by the LGS WFS that has 16x16 subapertures. Each subaperture has 2x2 2.5" pixels. The number of photodetections for the LGS WFS is about 100 photon/subaperture/exposure. The NGS WFS has variable number of subapertures given there are different sized lenslet arrays with the same focal length that are easily configurable in the instrument. There are 5 NGS WFS configurations: 16x16, 8x8, 5x5, 2x2, and 1x1. Assumptions made for the NGS in Section \ref{subsec:oldParams} still hold reasonable to the as-built Robo-AO-2, so the photometric parameters for the NGS are kept the same. Both wavefront sensors use NuVu HNu128AO EMCCD cameras for high speed wavefront sensing with detectors that are UV optimized in case they need to be used interchangeably. The noise parameters of these cameras and other relevant simulation parameters are tabulated in Table \ref{tab:noiseParamsHAPA}.
\begin{table}[h]
    \centering
    \begin{tabular}{|l|r|r|}
        \hline
        Parameter & LGS & NGS \\
        \hline\hline
        Central wavelength & 355 nm & 750 nm \\ 
        \hline
        Guide star altitude & 10 km& Infinity\\
        \hline
        Guide star intrinsic size & 1.2" & \\
        \hline
        Photometric zero point & & 6.9x10$^{10}$\\
        \hline
        Sky background & & 18.5\\ 
        \hline 
        Quantum efficiency & 0.68& 0.67\\
        \hline
        Optical throughput & 0.34& 0.30\\
        \hline
        Excess noise factor & 1.41 & 1.41\\
        \hline
        Read-out noise & 0.39 e$^-$ & 0.39 e$^-$\\
        \hline
        Dark current & 0.01 e$^-$/s & 0.01 e$^-$/s\\
        \hline
    \end{tabular}
    \caption{Photometry and noise parameters used in the simulations.}
    \label{tab:noiseParamsHAPA}
\end{table}

The common-path DM seen by the LGS WFS, NGS WFS, and the science field has 17x17 actuators that are registered to the LGS WFS and the NGS WFS in the 16x16 configuration using Fried geometry\cite{Fried1977}. The LGS has an uplink tip-tilt mirror inside the laser projector driven with a gain of 0.3 to stabilize the laser beam in the wavefront sensor camera. The NGS WFS and the science field has a separate tip-tilt mirror inside the instrument unseen by the LGS WFS. 

The minimum-variance tomographic reconstructor is again used for the HAPA simulations. The LGS-only results come from running HAPA simulations with the NGS WFS set to the 1x1 subaperture configuration, meaning all higher order correction depends on the LGS, the NGS is only used as a tip-tilt guide star. For the NGS simulations, a zonal regularized least-squares reconstructor was used. 

The loop rate we run the AO system is 1500 Hz, which is the frame rate we always run the LGS WFS at since the brightness of an LGS does not change dramatically during observations. The NGS WFS may operate with integration cycle choices of [1, 2, 4, 8, 16, 32, 64] cycles, which correspond to frame rates of [1500, 750, 375, 188, 94, 47, 23] Hz, depending on the magnitude of the stellar object used for guiding. All simulations use a loop gain of 0.4. Based on the results in Section \ref{subsec:oldResults} that showed the number of iterations do not significantly affect the end results, we set the simulations to run 3000 iterations per simulation, which is 2 s of integration time on-sky. The first 200 iterations are skipped for the AO loop to stabilize before statistics are starting to be calculated.

We ran all the simulations with the same typical turbulence profile (Table \ref{tab:50turbDome}). Compared to the turbulence profile in Table \ref{tab:50turb}, the only difference is that dome seeing of 0.4" is added and the turbulence fractions are adjusted accordingly. The r$_0$ parameter at 500 nm is also adjusted from 0.128 m to 0.113 m. The dome seeing in UH2.2 is the strongest at about 0.1 Hz, thus giving it a wind speed of 0.0113 m/s combined with r$_0$. The characteristics of dome seeing at UH2.2 come from Chun and Baranec, both of whom have had ample experiences observing with UH2.2.

\begin{table}[h]
    \centering
    \setlength{\tabcolsep}{5pt}
    \begin{tabular}{|l||r|r|r|r|r|r|r|r|r|r|r|}
        \hline
        Elevation (m) & 0 & 0 & 5 & 15 & 37 & 197 & 357 & 804 & 3171 & 7307 & 14636 \\
        \hline
        Turbulence fraction &0.183& 0.204 & 0.298 & 0.080 & 0.048 & 0.018 & 0.005 & 0.037 & 0.034 & 0.054 & 0.039\\
        \hline
        Wind speed (m/s) & 0.011&6.8 & 6.8 & 6.8 & 6.8 & 6.8 & 6.8 & 6.9 & 8.2 & 21.2 & 27.6\\
        \hline
        Wind direction (deg) & 48 & 0 & 76 & 92 & 190 & 255 & 270 & 350 & 17 & 29 & 66\\
        \hline
    \end{tabular}
    \caption{Typical turbulence profile used in the simulations. Different from Table \ref{tab:50turb} because 0.4" of seeing is added.}
    \label{tab:50turbDome}
\end{table}

\begin{figure}
    \centering
    \includegraphics[width=0.8\linewidth]{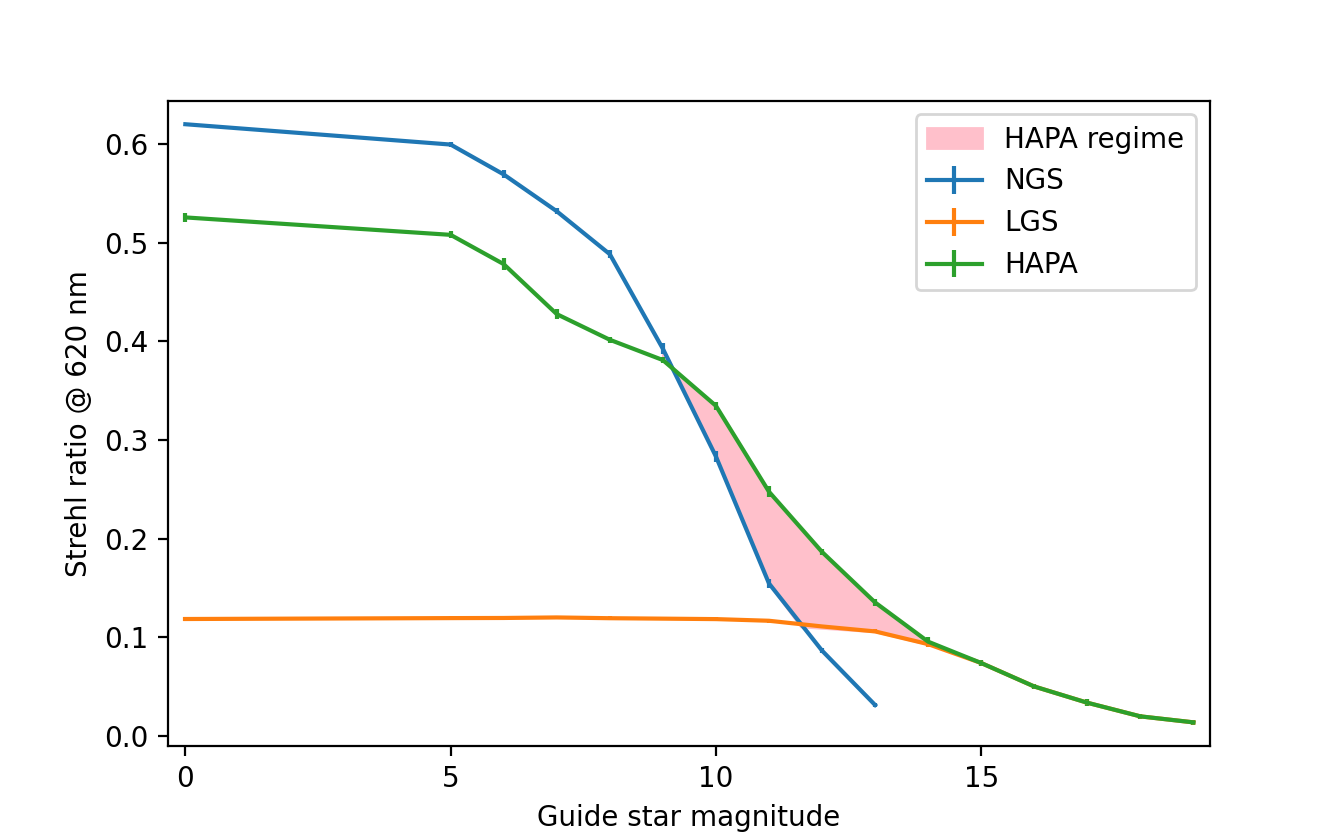}
    \caption{Simulated performances of NGS-only, LGS-only, and HAPA with the as-built Robo-AO-2. The ``HAPA regime" defines the Strehl ratio versus guide star magnitude space where HAPA outperforms both NGS and LGS-only modes.}
    \label{fig:HAPARobo-AO-2}
\end{figure}

\subsection{Simulation results}
\label{subsec:simResultsRobo-AO-2}
For each natural guide star magnitude, we ran simulations with the different NGS WFS configurations and different frame rates to optimize the resulting Strehl ratios. After finding the ideal NGS WFS configuration and frame rate for each NGS magnitude, simulations with the same ideal parameters were repeated 30 times and their medians were taken as the results (Table \ref{tab:newResults}; Figure \ref{fig:HAPARobo-AO-2}).

\begin{table}[h]
    \centering
    \scriptsize
    \setlength{\tabcolsep}{4pt}
    \begin{tabular}{|c||r|r|r|r|r|r|r|r|r|r|r|r|r|r|r|r|}
         \hline
         Mag&0&5&6&7&8&9&10&11&12&13&14&15&16&17&18&19 \\
         \hline
         \hline
         HAPA &0.526&0.508&0.478&0.428&0.402&0.381&0.335&0.248&0.186&0.135&0.096&0.074&0.05 &0.034&0.02 &0.014\\
         $\pm$STD&0.004&0.004&0.006&0.005&0.003&0.003&0.004&0.005&0.003&0.003&0.005&0.003& 0.003&0.004&0.002&0.002\\
         \hline
         NGS config& 16x16&16x16&16x16&16x16&5x5&5x5&5x5&5x5&2x2&2x2&2x2&1x1&1x1&1x1&1x1&1x1\\
         \hline
         NGS FR & 1500&1500&1500&750&750&750&750&375&375&188&188&94&94&47&23&23\\
         \hline
         \hline
         LGS-only&0.119&0.119&0.120&0.120&0.119&0.119&0.118&0.117&0.111&0.106&0.093&0.074&0.050 &0.034&0.020&0.014\\
         $\pm$STD & 0.002&0.002&0.001&0.002&0.002&0.002&0.002&0.002&0.002&0.002&0.003&0.003&0.003&0.004&0.002&0.002\\
         \hline
         NGS config&1x1&1x1&1x1&1x1&1x1&1x1&1x1&1x1&1x1&1x1&1x1&1x1&1x1&1x1&1x1&1x1\\
         \hline
         NGS FR & 750&750&375&375&375&375&375&375&188&188&188&94&94&47&23&23\\
         \hline
         \hline
         NGS-only & 0.620&0.599&0.569&0.532&0.488&0.393&0.283&0.155&0.086&0.031&&&&&&\\
         $\pm$STD & 0.000&0.002&0.004&0.003&0.004&0.006&0.005&0.005&0.003&0.002&&&&&&\\
         \hline
         NGS config& 16x16&16x16&16x16&16x16&16x16&16x16&16x16&16x16&8x8&8x8&&&&&&\\
         \hline
         NGS FR & 1500&1500&1500&750&750&750&375&375&375&375&&&&&&\\
         \hline
    \end{tabular}
    \caption{Optimal simulation results for HAPA, LGS-only, and NGS-only mode for the as-built Robo-AO-2 at each natural guide star magnitude with the corresponding NGS configuration and frame rate.}
    \label{tab:newResults}
\end{table}

Compared to the simulation results in Section \ref{subsec:oldResults}, the results for the as-built Robo-AO-2 have less magnitude versus Strehl ratio regime where HAPA outperforms the traditional NGS-only and LGS-only wavefront sensing modes. This is because in the initial set of simulations, the HAPA mode was simulated with a higher order system of 19x19 subapertures, whereas the NGS-only mode just had a WFS with 16x16 subapertures. In the results of the as-built Robo-AO-2, the NGS-only mode achieves a higher Strehl ratio than HAPA because HAPA is using a tomographic reconstructor that combines information coming from both the NGS WFS and the LGS WFS. Information provided by the LGS WFS has a dominating source of error, focal anisoplanatism, thus degrading the performance of HAPA compared to the NGS-only mode in the bright stars regime. As the natural guide star gets dimmer, the NGS curve quickly rolls off as the measurement noise drastically increases due to not detecting enough photons. The HAPA curve rolls off slower since the LGS WFS is still providing information for higher order correction and a lower-order NGS WFS is used to lessen the effect of focal anisoplanatism. Based on the Robo-AO-2 simulations, HAPA outperforms both NGS-only and LGS-only mode for guide star magnitudes ranging from 9 to 14. We will carry out corresponding on-sky observations to understand the validity and the accuracy of the simulations.

\section{On-sky experiment plans}
\label{sec:expPlans}
Currently, Robo-AO-2 is operating with the LGS WFS only for AO correction, which is its typical observing mode. The NGS WFS system in Robo-AO-2 has not yet been commissioned. Even though all the mechanical and optical components of the NGS WFS are assembled and aligned inside the instrument, they have not been integrated with the RTC and tested on-sky. The next step in this project is to commission the NGS WFS. We also need to develop the algorithms to drive the LGS and NGS WFSs together using tomographic reconstructors. Ideally, we aim to design the software such that switching between the different AO observing modes would be easy and instantaneous. Conducting observations for all three modes and the different guide star magnitudes in a short period of time is important because seeing conditions are ever changing. The chance for weather to drastically change during a few hours in the night is much smaller than the weather changing over a time span of multiple nights. A dataset good for comparison needs to be taken under similar weather conditions since weather can strongly affect the performance of an AO system.

Once the HAPA and the NGS-only observing modes operate on-sky, we will conduct a series of observations under various seeing conditions. Similar to the simulations, the effect of the NGS WFS configuration and the NGS frame rate will be examined to optimize the AO correction at each guide star magnitude for each observing mode. As we track a bright star across the sky and use it as the NGS, we would take data with it using all three observing modes, while changing its brightness by switching the ND filter in front of the NGS WFS. After a complete set of on-sky data is acquired, we would compare the on-sky results with the simulation results. This comparison should give us insights on what assumptions we are missing in our simulations and how we can polish them. We intend to use our experiences for implementing HAPA with Robo-AO-2 and its simulations to build more accurate simulations for extreme AO systems. These simulations would help the community understand how beneficial HAPA may be for the existing and future exoplanet direct imaging systems. The timeline for the project is tabulated below (Table \ref{tab:timeline}).

\begin{table}[h!]
    \centering
    \begin{tabular}{|l|l|}
         \hline
         2024 Summer& Commission the NGS WFS system in Robo-AO-2 \\
        \hline
         2024 Fall& Commission the HAPA observing mode in Robo-AO-2 \\
                  & Start taking on-sky data\\
        \hline
         2025 Spring& Continuing taking on-sky data \\
                    & Build simulations for exoplanet direct imaging systems\\
        \hline
    \end{tabular}
    \caption{Timeline for the HAPA project.}
    \label{tab:timeline}
\end{table}

\section{Conclusion and discussions}
\label{sec:conclusion}
Direct imaging of exoplanets can vastly improve our understanding of the worlds outside of our solar system. In most cases, extreme AO correction is required to conduct these observations, which means only stellar systems with a very bright guide star could be observed. There are only a limited number of systems with bright natural guide stars. To increase the number of observable targets, we propose a wavefront sensing technique called HAPA, which stands for Hybrid Atmospheric Phase Analysis. HAPA combines signals from a LGS WFS and a NGS WFS using a tomographic reconstructor; it relies on the LGS for higher order correction and uses the lower order NGS to allow high Strehl ratio correction using fainter natural guide stars. 

Robo-AO-2 serves as a perfect apparatus to carry out HAPA experiments and demonstrations given its many designed features. It has a LGS WFS that uses a UV Raleigh laser creating an artificial guide star at 355 nm and a NGS WFS. Both WFSs have variable frame rates that can be optimized depending on the guide star magnitude. A ND filter-wheel sits in front of the NGS WFS to easily change the guide star magnitude of the NGS. The NGS WFS also has different subaperture configurations of [16x16, 8x8, 5x5, 4x4, 2x2, 1x1] to optimize its signal to noise ratio and AO correction depending on the NGS brightness. A common path DM of 17x17 actuators is shared by both of the WFSs. Compared to large observatories, we are able to get an abundant of observing time under various weather conditions with UH2.2. 

Even though Robo-AO-2 is not a high order AO system, simulations have shown that there is still a regime where HAPA outperforms both the NGS-only and the LGS-only modes. After conducting on-sky experiments with Robo-AO-2 regarding its NGS, LGS and HAPA observing modes, we will use the lessons we learn to study the benefits HAPA may bring to existing and future extreme AO/exoplanet direct imaging systems.
\acknowledgments 
 
This work is partially supported by the National Science Foundation under Grant No. AST-1712014. 

\bibliography{report} 

\begin{thebibliography}{10}

\bibitem{Mayor1995}
{Mayor}, M. and {Queloz}, D., ``{A Jupiter-mass companion to a solar-type star},'' {\em Nature}~{\bf 378},  355--359 (Nov. 1995).

\bibitem{Davies2012}
{Davies}, R. and {Kasper}, M., ``{Adaptive Optics for Astronomy},'' {\em Annual Review of Astronomy and Astrophysics}~{\bf 50},  305--351 (Sept. 2012).

\bibitem{Martinache2009}
{Martinache}, F., {Guyon}, O., {Lozi}, J., {Garrel}, V., {Blain}, C., and {Sivo}, G., ``{The Subaru Coronagraphic Extreme AO Project},'' in [{\em Exoplanets and Disks: Their Formation and Diversity}{\nolinebreak\hspace{0.1em}]},  {Usuda}, T., {Tamura}, M., and {Ishii}, M., eds., {\em American Institute of Physics Conference Series} {\bf 1158},  319--322 (Aug. 2009).

\bibitem{Dekany2013}
{Dekany}, R., {Roberts}, J., {Burruss}, R., {Bouchez}, A., {Truong}, T., {Baranec}, C., {Guiwits}, S., {Hale}, D., {Angione}, J., {Trinh}, T., {Zolkower}, J., {Shelton}, J.~C., {Palmer}, D., {Henning}, J., {Croner}, E., {Troy}, M., {McKenna}, D., {Tesch}, J., {Hildebrandt}, S., and {Milburn}, J., ``{PALM-3000: Exoplanet Adaptive Optics for the 5 m Hale Telescope},'' {\em The Astrophysical Journal}~{\bf 776},  130 (Oct. 2013).

\bibitem{Macintosh2014}
Macintosh, B., Graham, J.~R., Ingraham, P., Konopacky, Q., Marois, C., Perrin, M., Poyneer, L., Bauman, B., Barman, T., Burrows, A.~S., Cardwell, A., Chilcote, J., Rosa, R. J.~D., Dillon, D., Doyon, R., Dunn, J., Erikson, D., Fitzgerald, M.~P., Gavel, D., Goodsell, S., Hartung, M., Hibon, P., Kalas, P., Larkin, J., Maire, J., Marchis, F., Marley, M.~S., McBride, J., Millar-Blanchaer, M., Morzinski, K., Norton, A., Oppenheimer, B.~R., Palmer, D., Patience, J., Pueyo, L., Rantakyro, F., Sadakuni, N., Saddlemyer, L., Savransky, D., Serio, A., Soummer, R., Sivaramakrishnan, A., Song, I., Thomas, S., Wallace, J.~K., Wiktorowicz, S., and Wolff, S., ``First light of the gemini planet imager,'' {\em Proceedings of the National Academy of Sciences}~{\bf 111}(35),  12661--12666 (2014).

\bibitem{Beuzit2019}
{Beuzit}, J.~L., {Vigan}, A., {Mouillet}, D., {Dohlen}, K., {Gratton}, R., {Boccaletti}, A., {Sauvage}, J.~F., {Schmid}, H.~M., {Langlois}, M., {Petit}, C., {Baruffolo}, A., {Feldt}, M., {Milli}, J., {Wahhaj}, Z., {Abe}, L., {Anselmi}, U., {Antichi}, J., {Barette}, R., {Baudrand}, J., {Baudoz}, P., {Bazzon}, A., {Bernardi}, P., {Blanchard}, P., {Brast}, R., {Bruno}, P., {Buey}, T., {Carbillet}, M., {Carle}, M., {Cascone}, E., {Chapron}, F., {Charton}, J., {Chauvin}, G., {Claudi}, R., {Costille}, A., {De Caprio}, V., {de Boer}, J., {Delboulb{\'e}}, A., {Desidera}, S., {Dominik}, C., {Downing}, M., {Dupuis}, O., {Fabron}, C., {Fantinel}, D., {Farisato}, G., {Feautrier}, P., {Fedrigo}, E., {Fusco}, T., {Gigan}, P., {Ginski}, C., {Girard}, J., {Giro}, E., {Gisler}, D., {Gluck}, L., {Gry}, C., {Henning}, T., {Hubin}, N., {Hugot}, E., {Incorvaia}, S., {Jaquet}, M., {Kasper}, M., {Lagadec}, E., {Lagrange}, A.~M., {Le Coroller}, H., {Le Mignant}, D., {Le Ruyet}, B., {Lessio}, G., {Lizon}, J.~L., {Llored}, M.,
  {Lundin}, L., {Madec}, F., {Magnard}, Y., {Marteaud}, M., {Martinez}, P., {Maurel}, D., {M{\'e}nard}, F., {Mesa}, D., {M{\"o}ller-Nilsson}, O., {Moulin}, T., {Moutou}, C., {Orign{\'e}}, A., {Parisot}, J., {Pavlov}, A., {Perret}, D., {Pragt}, J., {Puget}, P., {Rabou}, P., {Ramos}, J., {Reess}, J.~M., {Rigal}, F., {Rochat}, S., {Roelfsema}, R., {Rousset}, G., {Roux}, A., {Saisse}, M., {Salasnich}, B., {Santambrogio}, E., {Scuderi}, S., {Segransan}, D., {Sevin}, A., {Siebenmorgen}, R., {Soenke}, C., {Stadler}, E., {Suarez}, M., {Tiph{\`e}ne}, D., {Turatto}, M., {Udry}, S., {Vakili}, F., {Waters}, L.~B.~F.~M., {Weber}, L., {Wildi}, F., {Zins}, G., and {Zurlo}, A., ``{SPHERE: the exoplanet imager for the Very Large Telescope},'' {\em Astronomy and Astrophysics}~{\bf 631},  A155 (Nov. 2019).

\bibitem{hapaHawaiian}
``Hawaiian dicitonaries.'' \url{https://wehewehe.org/gsdl2.85/cgi-bin/hdict?d=D3021&l=en&e=d-11000-00---off-0hdict--00-1----0-10-0---0---0direct-10-ED--4--textpukuielbert%2Ctextmamaka-----0-1l--11-haw-Zz-1---Zz-1-home-hapa--00-3-1-00-0--4----0-0-11-00-0utfZz-8-00}.
\newblock Acceesed: 2024-04-22.

\bibitem{Ellerbroek1994}
Ellerbroek, B.~L., ``First-order performance evaluation of adaptive-optics systems for atmospheric-turbulence compensation in extended-field-of-view astronomical telescopes,'' {\em J. Opt. Soc. Am. A}~{\bf 11},  783--805 (Feb 1994).

\bibitem{Ellerbroek1997}
{Ellerbroek}, B.~L. and {Rhoadarmer}, T.~A., ``{Optimizing the performance of closed-loop adaptive-optics control systems on the basis of experimentally measured performance data.},'' {\em Journal of the Optical Society of America A}~{\bf 14},  1975--1987 (Aug. 1997).

\bibitem{Wizinowich2022}
{Wizinowich}, P., {Lu}, J., {Cetre}, S., {Chin}, J.~C., {Correia}, C.~M., {Delorme}, J.~R., {Gers}, L., {Lilley}, S.~J., {Lyke}, J.~E., {Marin}, E., {Ragland}, S., {Richards}, P., {Surendran}, A., {Wetherell}, E., {Chen}, G. C.-F., {Chu}, D., {Do}, T., {Fassnacht}, C., {Freeman}, M., {Gautam}, A.~K., {Ghez}, A., {Hunter}, L., {Jones}, T.~A., {Liu}, M., {Mawet}, D., {Max}, C., {Morris}, M., {Phillips}, M., {Ruffio}, J.-B., {Rundquist}, N.-E., {Sabhlok}, S., {Terry}, S., {Treu}, T., and {Wright}, S., ``Keck all sky precision adaptive optics program overview,'' {\em SPIE Astronomical Telescopes + Instrumentation}~{\bf 12185} (2022).

\bibitem{Akiyama2020}
Akiyama, M., Minowa, Y., Ono, Y., Terao, K., Ogane, H., Oomoto, K., Iizuka, Y., Oya, S., Mieda, E., and Yamamuro, T., ``Ultimate-start: Subaru tomography adaptive optics research experiment project overview,'' in [{\em Adaptive Optics Systems VII}{\nolinebreak\hspace{0.1em}]},  Schreiber, L., Schmidt, D., and Vernet, E., eds., {\em Proceedings of SPIE - The International Society for Optical Engineering}, SPIE (2020).
\newblock Funding Information: The ULTIMATE-START project is supported by the Japan Society for the Promotion of Science (Grant-in-Aid for Research No.17H06129). Publisher Copyright: {\textcopyright} 2020 SPIE.; Adaptive Optics Systems VII 2020 ; Conference date: 14-12-2020 Through 22-12-2020.

\bibitem{Baranec2014a}
{Baranec}, C., {Dekany}, R.~G., {Burruss}, R.~S., {Bowler}, B.~P., {van Dam}, M., {Riddle}, R., {Shelton}, J.~C., {Truong}, T., {Roberts}, J., {Milburn}, J., and {Tesch}, J., ``{PULSE: The Palomar Ultraviolet Laser for the Study of Exoplanets},'' in [{\em Adaptive Optics Systems IV}{\nolinebreak\hspace{0.1em}]},  {Marchetti}, E., {Close}, L.~M., and {Vran}, J.-P., eds., {\em Society of Photo-Optical Instrumentation Engineers (SPIE) Conference Series} {\bf 9148},  91481D (July 2014).

\bibitem{TRILEGAL}
{Vanhollebeke}, E., {Groenewegen}, M.~A.~T., and {Girardi}, L., ``{Stellar populations in the Galactic bulge. Modelling the Galactic bulge with TRILEGAL},'' {\em Astronomy and Astrophysics}~{\bf 498},  95--107 (Apr. 2009).

\bibitem{Baranec2018}
{Baranec}, C., {Chun}, M., {Hall}, D., {Connelley}, M., {Hodapp}, K., {Huber}, D., {Liu}, M., {Magnier}, E., {Meech}, K., {Takamiya}, M., {Griffiths}, R., {Riddle}, R., {Dekany}, R., {Kasliwal}, M., {Lau}, R., {Law}, N.~M., {Guyon}, O., {de Pater}, I., {Wong}, M., {Ofek}, E., {Hammel}, H., {Kuchner}, M., {Simon}, A., {Moore}, A., {Kissler-Patig}, M., and {van Dam}, M.~A., ``{The Robo-AO-2 facility for rapid visible/near-infrared AO imaging and the demonstration of hybrid techniques},'' in [{\em Adaptive Optics Systems VI}{\nolinebreak\hspace{0.1em}]},  {Close}, L.~M., {Schreiber}, L., and {Schmidt}, D., eds., {\em Society of Photo-Optical Instrumentation Engineers (SPIE) Conference Series} {\bf 10703},  1070327 (July 2018).

\bibitem{vanDam2016}
{van Dam}, M. and {Baranec}, C., ``{Simulations of PULSE on Mauna Kea},'' tech. rep., Flat Wavefronts Internal Report, Christchurch, New Zealand (Sept. 2016).

\bibitem{rigaut1a2013simulating}
Rigaut, F. and Van~Dam, M., ``Simulating astronomical adaptive optics systems using yao,'' (2013).

\bibitem{vanDam2021}
van Dam, M.~A.,  [{\em Wavefront Reconstruction and Control}{\nolinebreak\hspace{0.1em}]}, ch.~Chapter 11,  205--223, World Scientific (2021).

\bibitem{Baranec2024}
{Baranec}, C., {Ou}, J., {Riddle}, R.~L., {Zhang}, R., {McKay}, L., {Rampy}, R., {Bonnet}, M., {Hamilton}, I., {Ching}, G., {Young}, J., {Salama}, M., {Barnes}, P., {Jacobson}, S., {Onaka}, P., {Chun}, M.~R., {Powell}, K., and {Werber}, Z., ``{Commissioning results from the Robo-AO-2 facility for rapid visible/near-infrared AO imaging},'' International Society for Optics and Photonics, SPIE (2024).

\bibitem{Fried1977}
{Fried}, D.~L., ``{Least-squares fitting a wave-front distortion estimate to an array of phase-difference measurements},'' {\em Journal of the Optical Society of America (1917-1983)}~{\bf 67},  370 (Mar. 1977).

\end{thebibliography}
\bibliographystyle{spiebib} 

\end{document}